\documentclass[preprint,showpacs,preprintnumbers,amsmath,amssymb]{revtex4}

\usepackage{amssymb}
\usepackage{graphics}
\usepackage{epsfig}
\usepackage{graphicx}    % Include figure files
\usepackage{dcolumn}     % Align table columns on decimal point
\usepackage{amsthm}
\usepackage{bm}          % bold math
\usepackage{amsbsy}
\begin{document}
\title{Accurate evaluation of the Green's function of disordered graphenes}
\author{W. Zhu$^{1,2}$, Q. W. Shi$^{1,2 \dagger}$,
X. R. Wang$^{2,3 \ast}$, X. P. Wang$^1$, J. L. Yang$^{1}$, J. Chen$^{4,5}$, J. G. Hou$^{1}$}
\address{$^1$Hefei National Laboratory for Physical Sciences at
Microscale, University of Science and Technology of China, Hefei
 230026, China}
\address{$^2$Department of Physics, The Hong Kong University of
Science and Technology, Clear Water Bay, Kowloon, Hong Kong}
\address{$^3$School of Physics, Shangdong University,
Jinan, P. R. China}
\address{$^4$Electrical and Computer Engineering, University of
Alberta, Alberta, Canada T6G 2V4}
\address{$^5$National Research Council/National Institute for
Nanotechnology, Alberta, Canada T6G 2M9}
\email[Electronic
address:]{phsqw@ustc.edu.cn;phxwan@ust.hk}

\date{\today}

\begin{abstract}
An accurate simulation of Green's function and self-energy
function of non-interacting electrons in disordered
graphenes are performed. Fundamental physical quantities such as
the elastic relaxation time $\tau_e$, the phase
velocity $v_p$, and the group velocity $v_g$ are evaluated.
New features around the Dirac point are revealed, showing
hints that multi-scattering induced hybridization of
Bloch states plays an important role in the vicinity of
the Dirac point.
\end{abstract}

\maketitle

\textit{Introduction.} ${\LARGE -}$ Graphene, a single layer of
graphite, has been intensively studied in recent years because of
many intriguing transport properties \cite{Nov,Zhang}. Examples
include minimum conductivity and linear carrier density dependence
of conductivity \cite{Geim, Sarma}. Electrons in an ideal graphene
are governed by the relativistic massless Dirac equation and
exhibit a linear dispersion relation in the vicinity of the Dirac
point and zero density of state at the Dirac point \cite{Castro,WZF}.
Among many relativistic effects, the Klein paradox \cite{Geim1} is
arguably one of the most important effects that differentiates Dirac
electrons \cite{yyzhang} from the Schrodinger
electrons in disordered systems.

A deep understanding of Dirac electrons in disordered
graphenes requires extended knowledge on the self-energy
function in order to extract such fundamental physical quantities
as the phase velocity $v_p$, the group velocity $v_g$, and
the elastic relaxation time $\tau_e$. However, it is known
\cite{Economou} that accurate and reliable calculations are
quite difficult and nontrivial. For this reason, various approximations
have been employed in different theoretical studies.
So far, almost all calculations \cite{Hwang,Katsnelson} concerning
electron properties in disordered graphenes were performed
without fully considering disorder effects.
The wave-nature of Dirac electrons is more pronounced near the
Dirac point because of very large electron wavelength there.
It is known that interference due to multi-scattering leads to
weak localization and the Anderson localization in conventional
disordered electron systems\cite{Lee}. Quantum interference
also plays important roles in coherent wave propagation through
quasi-random \cite{xrw} and  random media\cite{Rossum,Albada}.
As it will be shown below, interference-induced the hybridization
of Bloch states is essential in understanding the diffusion
properties near the Dirac point in realistic disordered
graphenes, although previous calculations \cite{Hwang,Katsnelson}
only captured the essential physics far away from the Dirac point.

In this Letter, we present a systematic method to exactly
calculate Green's function and the self-energy of large
size disordered graphenes.
We extract accurate $\tau_e$, $v_p$, and $v_g$ values from the spectral
function $A(\textbf{k},E)$ derived from the self-energy function.
In comparison with the results from the self-consistent Born
approximation (SCBA), it is found that $\tau_e$ is
overestimated by the SCBA in the strong disorder regime.
We show that both $v_p$ and $v_g$ deviate significantly  from their
unrenormalized values and exhibit substantial energy dependence,
especially near the Dirac point. The effective group velocity is
larger than the effective phase velocity, but significantly lower than its
unrenormalized value when the mixing of different Bloch states is
dominant. Moreover, we generalize the Einstein relation to calculate
the conductivity of the disordered graphene.

\textit{Model-and-method.} ${\LARGE -}$ $\pi$-electrons of undoped
graphene can be modeled by a tight-binding Hamiltonian on a
honeycomb lattice of two sites per unit cell, $H_0 =t\sum\limits
_{<ij>}|i><j|+h.c.$, where $t=-2.7$ eV is the hopping energy.
The corresponding eigenvalues and eigenstates of $H_0$ near the
Dirac point are \cite{Castro,Ando}, respectively,
$E_{k,\pm}=\pm\hbar v_F^0 k$ and $|\textbf{k}\pm>=(|\textbf{k}A>
\pm e^{i\phi(\textbf{k})}|\textbf{k}B>)/2$, where $v_F^0$ is the
unrenormalized Fermi velocity, and $\hbar$ is the Plank constant.
$A$ and $B$ stand for $A$- and $B$-sublattices.
$\phi(\textbf{k})$ is the polar angle of the momentum
$\textbf{k}$, and $|\textbf{k}A(B)>=\frac{1}{\sqrt{N_{A(B)}}}\sum
\limits_{r_{A(B)}}e^{i\textbf{k}\cdot\textbf{r}_{A(B)}}|r_{A(B)}>$,
where $\textbf{r}_{A(B)}$ is the position vector of the $A(B)$-lattice
and $N_{A(B)}$ is the total $A(B)$-lattice points.
The plus (minus) sign denotes the conduction (valence) band.
The Green function of clean graphene is, in a diagonal basis,
$G_0(\textbf{k},E)=\frac{1}{E+i0^{+}-\hbar v_F^0 k}|\textbf{k}+><
\textbf{k}+|+\frac{1}{E+i0^{+}+\hbar v_F^0
k}|\textbf{k}-><\textbf{k}-|$.
A weak point-like disorder is introduced through
$V=\sum \epsilon_i |i><i|$ with $n_{imp}$
randomly distributed impurity sites where the on-site energy
$\epsilon_i$ of each impurity can take $-V_0$ or $V_0$
(measured in the unit of $t$) with equal probability.
A dimensionless parameter $\alpha=\frac{n_{imp}V_0^2A_c}{2\pi
(\hbar v_F^0)^2}$ can be used to characterize disorder strength.
Here, $A_c$ is the area of the unit cell. Our calculation shows
that the physical quantities such as self-energy are only
determined by the parameter $\alpha$\cite{anderson}.

The ensemble-averaged Green function is defined as
$G(\textbf{k}\pm,E)=\overline
{<\textbf{k}\pm|\frac{1}{E+i\eta-H_0-V}|\textbf{k}\pm>}$,
where the bar means the ensemble average.
It can be calculated by using the well developed Lanczos
recursive method\cite{Lancozs,kspace}. In order to obtain an
accurate ensemble-averaged Green's function near the Dirac
point with high energy resolution\cite{parameter}, a large
sample containing $N=L_x\times L_y\simeq 6.0$ millions
carbon atoms (2400$\times$2400) is used in our simulation,
where $L_x(L_y)$ is the number of atoms in $x(y)$ directions.
The periodic boundary condition is used in our simulation in
order to reduce the finite size effect. Thus, the wave vectors
are $k_x=n_x4\pi/3aL_x$ and $k_y=n_y4\pi/\sqrt{3}aL_y$, where
$n_{x(y)}$ is an integer and \textit{a} is the lattice constant.

\textit{Self-energy function.} ${\LARGE -}$ Fig. 1 shows the
calculated real (a) and imaginary (b) parts of the self-energy
function for $n_{imp}/N=10\%$ and $V_0=0.5$ (open squares)
and $2.0$ (open circles), respectively. The self-energy
function is defined as usual, where $\Sigma(\textbf{k},E)=G^{-1}_0
(\textbf{k},E)-G^{-1}(\textbf{k},E)$ \cite{Simon}.
In principle, the self-energy function depends on
energy $E$ and wave vector $\textbf{k}$.
However, our simulation finds that the self-energy function is not
sensitive to wave vector $\textbf{k}$, while the one-particle
Green's function depends on both $E$ and $\textbf{k}$.
This is not surprising since the scatterer size is much smaller
than the electron wavelength near the Dirac point so that the
inhomogeneous structure of disordered graphene can be
described well by an effective homogeneous medium.
This is also why the self-energy function is assumed to be
$\textbf{k}$-independent in many perturbative calculations.
Our finding validates this assumption\cite{Simon}.

Our calculation should be compared with widely used results from
the self-consistent Born approximation (SCBA) that predicts
\cite{Ando,Mirlin}
\begin{eqnarray}
\Sigma(E)=\{\begin{array}{ccc}
-2E/\alpha-i2\Gamma_0, \quad \qquad \qquad |E|\ll\Gamma_0\\
-2\alpha (E+i\pi\alpha|E|) ln|\frac{E_c}{E}|-i\pi\alpha|E|,
\quad |E|\gg\Gamma_0 \\
\end{array} \nonumber
\end{eqnarray}
where $\Gamma_0=E_c e^{-1/\alpha}$ ($E_c$ is the cut-off energy)
As shown in Fig. 1, SCBA results agree well with our
exact self-energy function for very weak disorder
($V_0$=0.5 and $n_{imp}/N$=10$\%$)\cite{strength}.
When the disorder strength increases several times ($V_0$=2.0 or
$\alpha$$\sim$0.07), the perturbative results can not capture
the main features, especially near the Dirac point.
The discrepancy is obvious for $Im\Sigma(0)$ as shown in Fig. 1(b).
$Im\Sigma(E)$ at the Dirac point is $\Gamma_0\sim$$10^{-5}t$
($E_c=\sqrt{3}t$ \cite{WZF}) from the SCBA, while our exact value is
$Im\Sigma(0)\sim$$10^{-2}t$. Thus, the true broadening of states at or
near the Dirac point is much greater than what has been predicted by SCBA.
This huge discrepancy can be attributed to the
mixture of Bloch states caused by the impurities. Furthermore, the
level-repulsion effect pushes all energy level toward the Dirac
point so that the density of states at the Dirac point increases more
in the presence of impurities. Thus, the impurities make the imaginary
part of the self-energy function (directly associated  with the density
of states) at the Dirac point bigger.
When the wavelength becomes short and the quantum interference as
well as the Bloch state mixing are less important, $Im\Sigma(E)$
are determined by the disorder scattering. The difference between
our exact simulation and that of the SCBA is small
as shown in Fig. 1.

\begin{figure}
\includegraphics[width=0.5\textwidth]{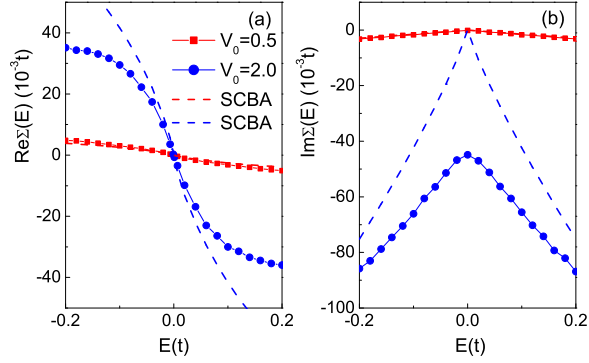}
\caption{(Color online) (a) Real part of self-energy as a function
of energy. (b)Imaginary part of self-energy as a function of energy.
The open squares and open circles represent our numerical calculations
for the disorder concentration $n_{imp}/N=10\%$ and $V_0=0.5$ and
$V_0=2.0$, respectively. Dashed lines represent the SCBA results
for the same disorder. Energy is measured in the units of $t$.}
\end{figure}

\textit{Spectral function.} ${\LARGE -}$ The single-particle
spectral function relates to Green's function through
$A(\textbf{k}\pm,E)=-Im G(\textbf{k}\pm,E)/\pi$. Fig. 2(a) is
$A(\textbf{k}+,E)$ for $k_y=0$, $V_0=1$, $n_{imp}/N$=10$\%$,
and various $k_x$ (curves from  the left to the right in the
figure) ranging from 0.0 ($n_x$=0) to 0.098 ($n_x$=56) (in unit
of $a^{-1}$). In the absence of disorder, the spectral function
$A_0(\textbf{k}\pm,E)$ is a delta function, reflecting that the
wave vector $\textbf{k}$ is a good quantum number and has
all its spectral weight precisely at the energy $E=E_{k\pm}$.
In the presence of disorder, the translational symmetry is broken
and the spectral function is broadened due to the
disorder scattering effect. The widths of the spectral function
are given by $Im\Sigma(E)$ that measures the elastic relaxation
lifetime $\tau_e$, where $\tau_e=\frac{\hbar}{-2Im\Sigma(E)}$.
Therefore, the elastic scattering relaxation time is akin
to $Im\Sigma(E)$, and $\tau_e$ around the Dirac point is mainly
determined by the Bloch state mixing and the level repulsion
effect\cite{Matteo}. Far away from the Dirac point,
the lifetime is mainly attributed to the disorder scattering.
As shown in the inset of Fig. 2, lifetime become shorter as the
wave vector increases. Physically, this is because the density
of states $\rho$ increases linearly with energy and the disorder
scattering effects become larger, which is qualitatively consistent
with the prediction of SCBA\cite{MacDonald,Ando,Mirlin}.

\begin{figure}
\includegraphics[width=0.45\textwidth]{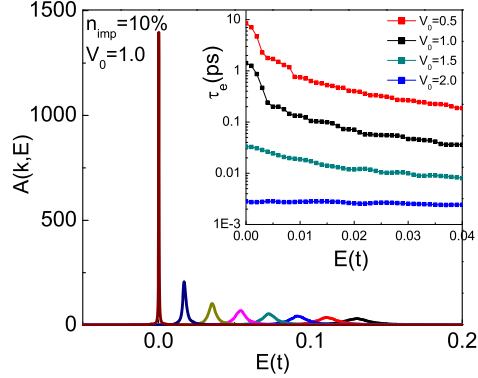}
\caption{(Color online) Single-particle spectral function
$A(\vec{k}+,E)$ plotted as a function of energy $E$ at several
k-points (from left to right): $k=0.000,\ 0.014,\ 0.028,\ 0.042,
\ 0.056,\ 0.070,\ 0.084,\ 0.098$ (or $n_x=0\sim56$) along the
$k_x$ direction. The model parameters are $n_{imp}/N=10\%$ and
$V_0=1.0$. Inset: The energy dependence of single-particle
relaxation time $\tau_e$ for $V_0=0.5\sim2.0$ and $n_{imp}/N=10\%$.}
\end{figure}

\textit{Effective band velocity.} ${\LARGE -}$ Dirac electron
propagation velocities, including the group velocity $v_g$ and phase
velocity $v_p$, are also greatly modified by the disorder effects.
These quantities relate to the shape of the dispersion relation that
is the roots of $E-E_0(k)-Re\Sigma(E)=0$\cite{Mahan,Kartick}.
One can also extract the dispersion relation from the peak of the
spectral function $A(\textbf{k},E)$ for a given $\textbf{k}$.
Fig 3 (a) shows our exact dissipation curve $E_{eff}(k)$ that
is linear for very weak disorder ($\alpha < 0.01$).
However, when the disorder strength becomes significantly large ($\alpha
\sim 0.1$), the $E_{eff}(k)$ becomes concave near the Dirac point,
indicating the reduction in both the group and phase velocities.

Fig. 3(b) shows $k-$dependence of the group and phase velocities obtained
from $v_g=\partial E_{eff}(k)/\partial k$ and
$v_p=E_{eff}/k$. For the very weak disorder $\alpha=0.01$ as
shown in the figure, $v_g$ and $v_p$ are not too
much different from the unrenomalized velocity $v_F^0$.
Their values are reduced by $5\%$ in comparison with $v_F^0$.
When the disorder strength increase several times $\alpha=0.07$,
the renormalized $v_g$ is higher than $v_p$ and both $v_g$ and
$v_p$ are reduced by a large percentage near the Dirac point.
This shows that disorder not only renormalizes the Dirac electron
velocities, but also changes the linear dispersion relation.
The fact of large reductions of velocities at the Dirac point
indicates that Dirac electrons near the Dirac point are more sensitive to the disorder.

One can use the renormalization factor $Z$ defined as
$Z=(1-\frac{\partial Re\Sigma(E,\textbf{k})}{\partial E})^{-1}$
\cite{Steven} to measure the effect of disorder on electronic
structure. Its value equals the ratio $v_g/v_F^0$. As shown in
Fig. 3(b), $Z$ is very close to $1.0$ for very weak disorder.
In this regime, the Bloch state is still a good starting point for
understanding disordered graphene, while the transport properties are
expected to be described by the quasi-classical Boltzmann theory.
When the disorder strength increases several times, calculations shows
that $Z$ is much smaller than $1.0$, especially around the Dirac point.
This unusual feature directly reflects that a Bloch states around
the Dirac point couples strongly with other nearby Bloch states.
Therefore, the Bloch states would not be a good starting point to
perform perturbative calculations. In fact, the similar feature has
been observed in semiconductor alloys and has been termed non-Bloch
nature of alloy states\cite{Wang}. This is why the SCBA method cannot
produce accurate enough results. This nontrivial feature
could lead to intriguing transport and optical properties.
For example, linear dispersion relation leads to $E_n=\pm v_F^0
\sqrt{2e\hbar nB}\ \ \ \ \ (n=0,1,2...)$ LL series. Since the LLs
come from the quench of electron's kinetic energies, lower group
velocity near the Dirac point means the energy gap between $n=0$
LL and $n=\pm 1$ LL should be substantially reduced by the disorders
while that between $n=\pm 1$ LLs and $n=\pm 2$ LLs will be affected
less by disorder. As a result, one should expect the ratio
$(E_2-E_{-1})/(E_1-E_0)$ is substantially larger than $\sqrt{2}+1$.
This is indeed what was observed in a recent experiment\cite{kim}
where $(E_2-E_{-1})/(E_1-E_0)$ is found to be greater than
$\sqrt{2}+1$, by $5\sim 10 \%$. This observation implies that their
samples are quite disordered.

\begin{figure}
\includegraphics[width=0.5\textwidth]{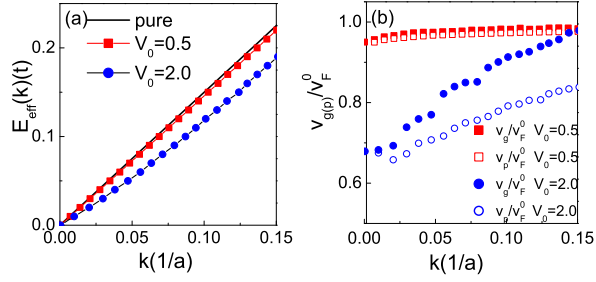}
\caption{(Color online) (a) Effective energy dispersion relation
for ideal graphene and disordered graphenes for $n_{imp}/N=10\%$
and $V=0.5$ (filled squares) and  $V=2.0$ (filled circles).
(b) The ratio $v_g/v_F^0$ (filled symbols) and $v_p/v_F^0$ (open
symbols) as a function of wavenumber $k$ (in units of $1/a$).}
\end{figure}

\textit{Conductivity.} ${\LARGE -}$ Assuming that the
Einstein relation for the Dirac electrons is valid, one can obtain the
conductivity from $\sigma=e^2 D \rho$, where $e$ is the electron
charge, $D=v_g^2\tau_{tr}/2$ is the diffusion coefficient, and
$\tau_{tr}$ denotes the transport relaxation time and satisfies the relationship
 $\tau_{tr}=2\tau_e$ for the point-like disorder. \cite{Ando}
Fig. 4 is the conductivity calculated from transport relaxation time
$\tau_{tr}=2\tau_e$ in Fig. 2, the group velocity $v_g$ in Fig. 3, and the
density of states obtained by $\rho(E)=-\frac{A_c}{\pi}
\int_0^{k_c}\frac{d\vec{k}}{(2\pi)^2}A(\vec{k},E)$ as shown in
inset to Fig. 4. For very weak disorder, the conductivity
is sub-linear in the Fermi energy.
Increasing the disorder strength, the conductivity weakly  depends
on the Fermi energy or the carrier density, in agreement with the prediction
of Boltzmann theory for short-range scatterers \cite{Ando}.
At the Dirac point, by combining the self-energy function with
Kubo formula, the conductivity is predicted to have
the universal value $\sigma_{xx}(0)=4e^2/\pi h$\cite{Yi}. However,
our calculations find that $\sigma_{xx}(0)$ takes the values $8.2e^2/ h$, $5.9e^2/ h$ and $2.7e^2/ h$
for $\alpha=0.0046$, $0.018$ and $0.07$, respectively, and shows non-universal behavior.
\begin{figure}
\includegraphics[width=0.45\textwidth]{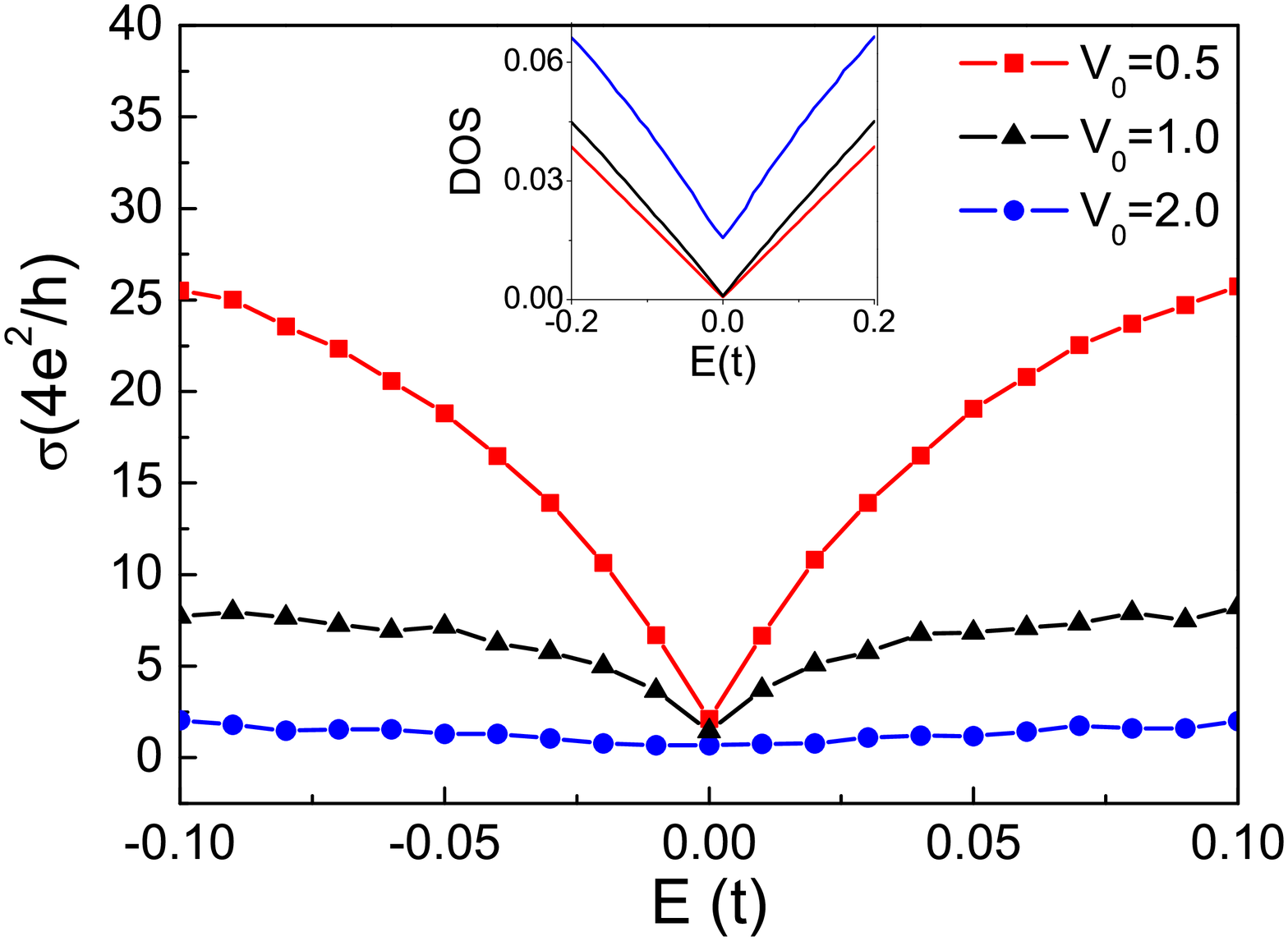}
\caption{(Color online) Conductivity as a function of charge density.
The model parameters are $n_{imp}/N=10\%$ and $V_0=$0.5, 1.0 and 2.0.
Inset: Density of states for the same parameters.}
\end{figure}

\textit{Summary.} ${\LARGE -}$ In conclusion, we studied point-like
disorder effects on the one-electron properties of graphene.
The exact ensemble-averaged Green's function is obtained from
a large-scale real-space calculation. Through the analysis of
self-energy and spectral functions, we conclude that the
single-particle lifetime reduction and the linear dispersion
relation are modified by the hybridization of the Bloch states.
Furthermore, we studied the diffusion transport properties by
using our exact self-energy and the Einstein relation.
Our approach is very general and robust, thus is applicable to
many other disordered systems.

\textit{Acknowledgment.}${\LARGE -}$ This work is partially supported
by NNSF of China (Nos. 10974187,10874165, and 50721091), by NKBRP of
China under Grant No. 2006CB922000, and
by KIP of the Chinese Academy of Sciences
(No. KJCX2-YW-W22). JC is supported by the NRC and NSERC of Canada
(No. 245680). XRW acknowledges the support of Hong Kong RGC grants
(\#604109, RPC07/08.SC03, and HKU10/CRF/08-HKUST17/CRF/08).

\end{document}